# NETWORKS UTILIZATION IMPROVEMENTS FOR SERVICE DISCOVERY PERFORMANCE

Intisar Al-Mejibli[1], Martin Colley[2] and Salah Al-Majeed[3]

[1, 2, 3] Department of Computer Science and Electronic Engineering, University of Essex, Colchester, UK
ialmej@essex.ac.uk
martin@essex.ac.uk
ssaleha@essex.ac.uk

## *ABSTRACT*

*Service discovery requests' messages have a vital role in sharing and locating resources in many of service discovery protocols. Sending more messages than a link can handle may cause congestion and loss of messages which dramatically influences the performance of these protocols. Re-send the lost messages result in latency and inefficiency in performing the tasks which user(s) require from the connected nodes. This issue become a serious problem in two cases: first, when the number of clients which performs a service discovery request is increasing, as this result in increasing in the number of sent discovery messages; second, when the network resources such as bandwidth capacity are consumed by other applications. These two cases lead to network congestion and loss of messages. This paper propose an algorithm to improve the services discovery protocols performance by separating each consecutive burst of messages with a specific period of time which calculated regarding the available network resources. It was tested when the routers were connected in two configurations; decentralised and centralised .In addition, this paper explains the impact of increasing the number of clients and the consumed network resources on the proposed algorithm.*

## *KEYWORDS*

*Dropped messages, Service discovery protocols & Network Utilization.*

## 1. INTRODUCTION

The home network has become recognized as the forefront of the networking revolution, where consumer technology and Internet infrastructure intersect to change the way we lead our lives. A fast growing increase in the uses of home networks has been noticed, for example recent research from Pike Research predicts a strong growth in the intelligent lighting control market. Global revenue is expected to increase from $1.3 billion to $2.6 billion by 2016 in intelligent lights [1].

In fact, home network would consist of everything a homeowner could imagine, ranging from large domestic appliances such as the fridges, microwaves, audio-visual equipment to the lightweight temperature and smoke sensors. In addition to mobile devices, smart cards, bar codes in grocery packages and little chips in clothing and accessories. The main goal of interconnecting the home devices together is to share the network services and resources, and to invoke them remotely. Many protocols have been proposed to achieve this purpose which is locate and invoke the services and resources in network known as services discovery protocols [2]. Most of the service discovery protocols rely on the exchange of messages to locate remote services and to provide access to them. Sending too many messages into the network from multiple nodes at the same time, could cause congestion which will lead to router queue overflow and the loss of messages. Accordingly, more messages must be sent to discover the services in the network and this causes more latency in discovery process and greedy consumption of the network resources.

This paper discusses how to avoid dropped messages during service discovery process in small networks which fall in (local Area Network) LAN category such as a networked office building, or home. In addition, it proposes an algorithm to overcome this significant issue, in order to make the discovery process perform smoothly and seamlessly. Further it explains the impact of increasing the number of clients and rate of consumed network resources on the proposed algorithm results.

This paper is structured as follows. Section 2 introduces the related work which includes service discovery protocols and available mechanisms and algorithms that have been proposed to avoid or minimize the number of dropped messages. Section 3 introduces the proposed algorithm. The simulation results are detailed in section 4. Finally, the conclusion and future work are given in section 5.

## 2. RELATED WORK

We should introduce service discovery protocols and the relevant work mechanism in order to understand the subsequent sections. Service discovery protocols acts vital role in sharing and locating resources in network and many of these protocols depend on messaging to achieve their process. From this point of view, it is a significant to introduce service discovery protocols to understand the proposed algorithm properly.

### 2.1 Service Discovery Protocol (UPnP)

Service discovery protocols enable devices to discover all services in a network and some of them allow devices that provide services to announce their services. Each service discovery protocol must have two components: a client which is the component that has a set of requirements that form the services it needs, and a device which is the component that offers its service(s) and is requested by client. Accordingly, any node in a network may be a client, a device, or a client and device at same time. Service discovery protocols can be classified into two types: Registry-based such as Jini [4][5] and Peer-to-Peer like UPnP [3]. Registry-based can be classified into centralized registry like Jini and distributed registry like Service Location Protocol (SLP) [4][6]. The Registry-based and Peer-to-Peer approaches both have advantages and drawbacks. For example: Registry-based is well organized and managed, but the registry node could cause a bottle neck problem for the entire network since if this node is damaged for any reason the clients are not able to access the required services. While in the Peer-to-Peer type all services send messages regularly even if there isn't a target client and this causes an unnecessary consumption for the networks' resources. Some protocols consider the announcement as an essential principle in service discovery issue such as UPnP whereas others protocols do not use the announcement approach such as Bluetooth [4] [7]. A selection technique should be used to select the most appropriate service when the discovery phase results in two or more identical services. There are two selection modes: manual and automatic modes. In manual mode, service selection is the responsibility of the user entirely. This mode has drawbacks: users may not know enough about the services to distinguish among them and too much user involvement causes inconvenience. This mode is applied in all the investigated service discovery protocols. In automatic mode, the service discovery protocol selects the service this simplifies client programs. On the other hand automatic selection may not be select the choice that user wants.

Each service discovery protocol has a specific features and philosophy which are different from other protocols. Here we will explain UPnP in more details as it is used in our simulated model.

UPnP is proposed for use in home and small office environments and targets device and service discovery. It has the capability of automatically assigning IP addresses to networked devices. The components considered in UPnP are control points (clients) which are optional and devices (offers service(s)). Service discovery in UPnP is depends on the Simple Service Discovery Protocol (SSDP) [5]. SSDP was proposed to discover devices and services in a network easily, quickly, dynamically, and without any a priori knowledge. It uses HTTP over unicast and multicast UDP packets to define two functions: search the services of a network and announce the availability of services in a network. UPnP cannot scale well since it uses multicasting extensively (multicasting is used both for service advertisements and service requests) [6]. When a control point is connected to network, it starts requesting the required service(s) by sending multicast message over UDP transport protocol. The service(s) that match the required criteria responds by sending unicast message to requested control point. Consequently, the control point gets information about the requested service. On the other hand, when the device is connected to network, it starts announcing its service(s) regularly by sending multicast message (over UDP). F 1 shows multicast M-SEARCH format of UPnP protocol.

```
M-SEARCH * HTTP/1.1
HOST: 239.255.255.250:1900
MAN: "ssdp:discover"
MX: seconds to delay response (MUST be greater than or equal to 1 and SHOULD be less than 5 inclusive)
ST: search target
USER-AGENT: OS/version UPnP/1.1 product/version
```

Figure 1. Multicast M-SEARCH format of UPnP protocol

MX field value contains maximum wait time in seconds. Device responses should be delayed a random duration between 0 and this many seconds in order to balance load for the control point (client) when it processes responses. The devices may assume an MX field value less than that specified in the MX header field. In another words, if the MX header field specifies a field value greater than 1, the device should assume that it contained the value 1 or less. Devices must not stop responding to other requests while waiting the random delay before sending a response.

## 2.3 Other Techniques

There are a number of techniques have been proposed to avoid dropping packets such as the approach which suggested by Parry and Gangatharan. The principle of their idea is the packet size of each source should be adjusted according to the network bandwidth to optimize the network utilization and also to avoid packet overflow at the client buffer. Their approach is based on a controller which is used to trace the data transmission rate at the router. When the total transmission rate is higher than the network bandwidth, the transmission controller adjusts the packet size of the source nodes so that the transmission rate is equal to the network bandwidth. [8].

Jacobson suggested an end-to-end congestion avoidance mechanism as used in Transmission Control Protocol (TCP). These mechanisms have worked well on low bandwidth delay product networks, while with newer high-bandwidth delay networks they have shown to be inefficient and prone to be unstable [9].

Jin proposed an alternative to the end-to-end congestion avoidance mechanism, named Network Lion and used in Transmission Control Protocol (TCP) too. Network Lion is developed as a part of a new network transmission protocol. His method uses packet drop avoidance (PDA) mechanism which is based on the maximum burst size (MBS) theory. In addition, he uses a real-time available bandwidth algorithm. Network Lion does redesign the transmission control, as well as separating the pacing control in layer 3 and retransmission control in layer 4. [10].

Kevin Mills and Christopher Dabrowski [11] proposed four Algorithms for adaptive-jitter control depending on network size, in order to minimize the dropping of messages from the message queues in the UPnP protocol. In fact, UPnP permits clients to include a jitter bound in multicast (M-Search) queries in order to limit implosion. Qualifying devices use the jitter bound to randomize timing of their responses. Kevin Mills and Christopher Dabrowski's algorithms depend on the principle of this bound. All four of these algorithms are based on making each root device independently estimate the time it will take for all root devices to respond to each M-Search query. Each root device then uses its estimate to determine a time to send its own responses (if any). Each response message includes a value recommending how long the control-point M-Search task should listen for responses, so M-Search task does not need to guess an appropriate required maximum time for listening.

All root devices must send and listen to Notify messages (which include a caching time or *max-age*). When all root devices receives these messages, they should build a map (*NM*) of devices and services in the network. Consequently, a root device could use its *NM* to estimate how many response messages will be sent by all root devices. They assume that the messages will be sent consecutively at rate R and root devices will send messages sequentially in the ascending order of their unique identities.

## 3. THE PROPOSED ALGORITHM

The aims of the proposed algorithm are: determines the required sending queue space and the required time for the routers to forwards all the messages of burst mode before receiving the next burst of messages. The proposed algorithm includes the instruction and equations that explain the relation between the required queue sizes and the interval separating two consecutive bursts of messages, to avoid dropping messages. Regarding the Open Systems Interconnection model OSI, the algorithm has designed to be applied in application layer. It could be included in the protocols and applications codes, which may cause a burst of messages to the network in their strategies.

The following rules must be applied to compute the sending queue size in each router or the space which required being available in the sending queue of each router at the sending time and calculate the best interval for each router. ), Two algorithms are proposed based on the network topology (Decentralised and Centralised, because the configuration of the network affects the follow of the messages.

### 3.1 Decentralised Algorithm

This algorithm is for Decentralised network topology which its routers connected in Decentralised method.

#### 3.1.1 Queue size Algorithm:

The Algorithm which is used to calculate the size of the sending queue for each router is illustrated in F 2. The values m and n represent the number of clients and services that connected to Ri respectively, where i=1, 2 … No. of routers.

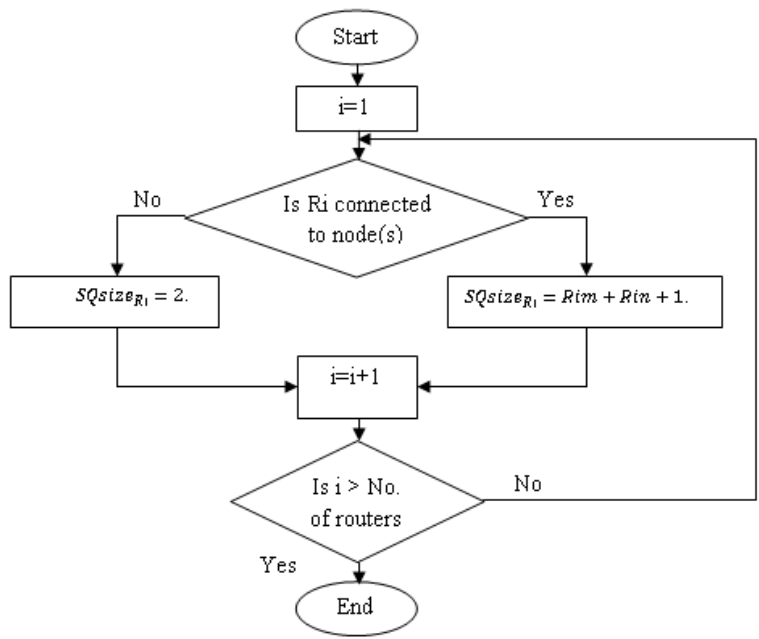

Figure 2. Flowchart of Queue Size Algorithm

### 3.1.2 Best Interval Algorithm:

The best interval algorithm is explained in F 3. In the suggested network topology any chosen router will divide the network into two parts, left and right. Equation (1) which is mentioned in F 3, guarantees that a specific router would forward all the receiving messages to their destination (client) before receiving the next burst of messages. It can be developed and take into consideration the available queue size for the specified router, as it represents the sharing space between all the clients (receivers) connected to that router so an overlap between two or more consecutive burst of messages can be achieved in order to minimize the required interval. Note z is the number of candidate routers.

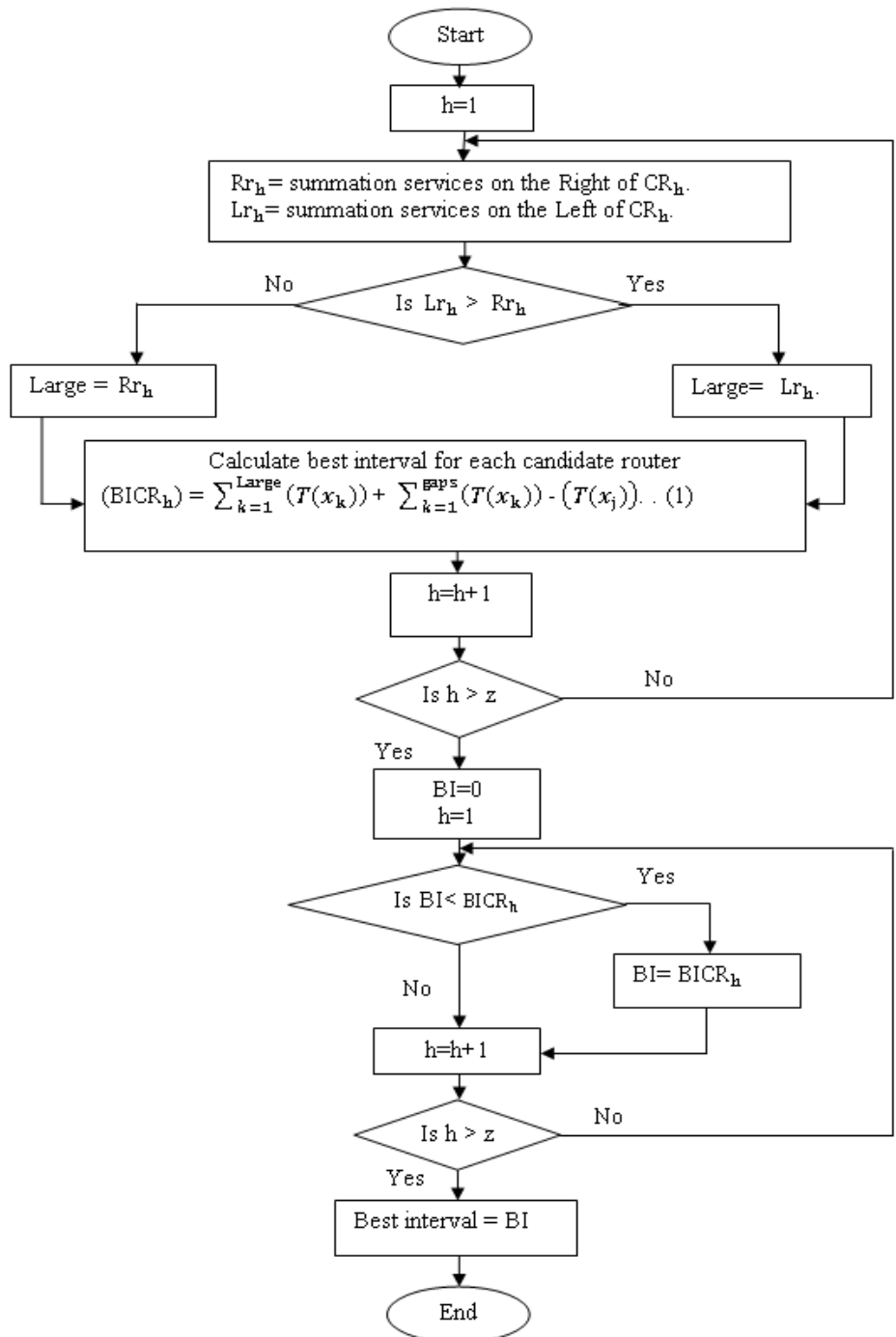

Figure 3. Flowchart of Best interval Algorithm

Where, $(T(x_k)) = \frac{\text{Message Size of service }_k}{\text{Bandwidth that message would use}}$

In Formula (1) the $(T(x_j))$ value is the biggest among $(Time(x_k))$, $k = 1,2,\ldots Large$. $(T(x_j))$ represents the time the message utilizes the link.

$\sum_{k=1}^{gaps}(T(x_k))$: represents the number of message times during which a specific router doesn't receive any service messages from nearest router(s). Here the average message size and average bandwidth is used. When there is a service connected directly to the nearest router, it would need at least two message times to reach the evaluated router.

The identified queue size in routers can be used to minimize the BI value. Overlapped space (OS) value represents this minimization. F 4 show how the Overlapped space (OS) is calculated:

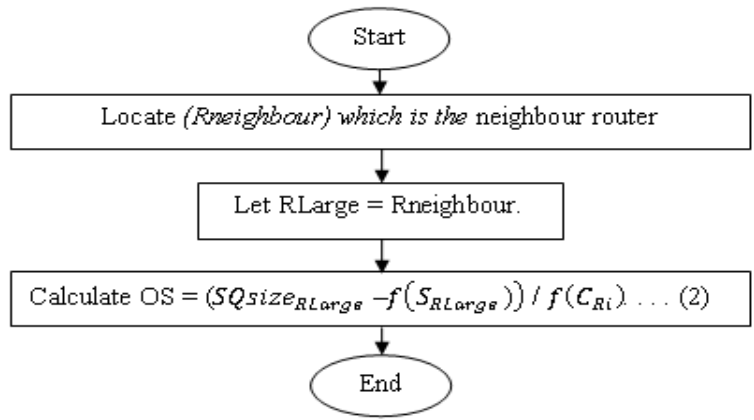

Figure4. Flowchart of OS Calculation Steps

Where $f(S_{RLarge}) = \sum_{k=1}^{n}(S_k)$ is number of all services that connected to the RLarge and

$f(C_{Ri}) = \sum_{k=1}^{m}(C_k)$ is number of all clients that connected to the and Ri. OS is measured by messages number and it represents the empty messages space in the sending queue of the chosen router. Equation (1) could use OS value and could be written as:

The best interval (BI) = $\sum_{k=1}^{Big} T(x_k) + \sum_{k=1}^{gaps} T(x_k) - T(x_j) - \sum_{k=1}^{OS} T(x_k)\ldots$ **(3)**

The question now, must each router in a network be evaluated in order to identify the best interval for entire network? And which interval would be used for the network? The answer is: Not all routers in a network must be evaluated instead some of them would be candidate to be evaluated and the longest interval will be used at the end, because, logically using the longest interval will avoid dropping messages at all other routers.

There are some conditions that help to identify which router will have the most impact in determining the best interval.

### 3.1.3 Choosing candidate router rules:

1- Identify the longest path between a service and a client. Then the router which connected to this client must be selected.

2- Identifying the router that connected to the largest number of clients and receives the largest number of services from one side of the network.

3- Identifying the router that connected to one or more clients and located nearest the end of the network.

4- If the chosen router is connected to one client then the nearest router connected to client too must be chosen, in order to compare between two consecutive burst of messages reach these routers consecutively.

In case that the two (or more) consecutive burst of messages were sent to the same client and this client is the lonely client connected to router, this means logically there are two (or more) receivers connected to that router and this should be taken into consideration in calculating the (OS) value.

One client may satisfy more than one of the previous conditions, in other words the client that has longest path with a service could be the same client that connected to a router which receives largest number of services and this wouldn't cause any problem.

All the candidate routers must be evaluated and the longest interval is the best interval for the network which would guarantee no losing messages.

### 3.2 Centralised Algorithm

This algorithm is proposed for the networks which its routers connected in centralised topology. All routers in this topology are connected to one router.

#### 3.2.1 Queue Size Algorithm:

F 5 illustrated the Algorithm which is used to calculate the size of the sending queue for each router. The values m and n represent the number of clients and services that connected to Ri respectively, where i=1, 2 ... No. of routers. RLargeS represents the router connected to the largest number of services and RLargeSni represents the number of services which connected to RLargeS.

#### 3.2.2 Best Interval Algorithm:

It is clear that any sending message between any two routers must pass the root router, in order to reach its destination. F 6 represent the best interval algorithm. Where RLargC[i], i=1, 2...j is an array of the router(s) that connected to the largest number of clients and RLargeC_S is the router that connected to the largest clients number and lowest services number.

Where $f(C_{Ri}) = \sum_{k=1}^{m}(C_k)$ is number of all clients that connected to the and Ri.

and $f(S_{Ri}) = \sum_{k=1}^{m}(S_k)$ is number of all services that connected to the and Ri.

Where, $(T(x_k)) = \frac{Message\ Size\ of\ service_k}{Bandwidth\ that\ message\ would\ use}$

$(T(x_j))$ value is the biggest among $(Time(x_k))$, $k = 1,2,...Large$. $(T(x_j))$ represents the time the message utilizes the link.

$\sum_{k=1}^{\text{gaps}}(T(x_k))$: represents the number of message times during which a specific router doesn't receive any service messages from nearest router(s). Here the average message size and average bandwidth is used.

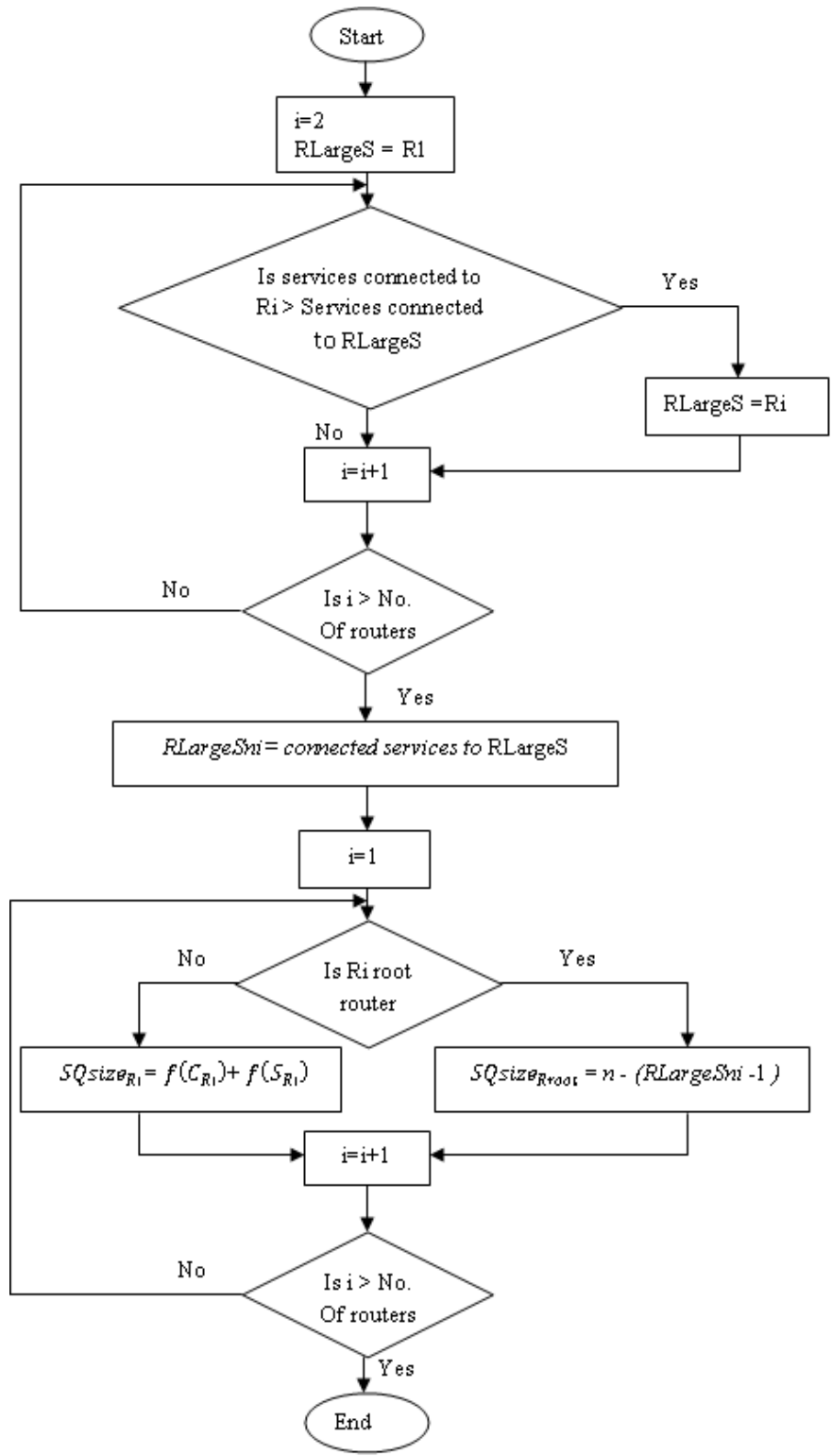

Figure 5. Flowchart of Queue Size Algorithm

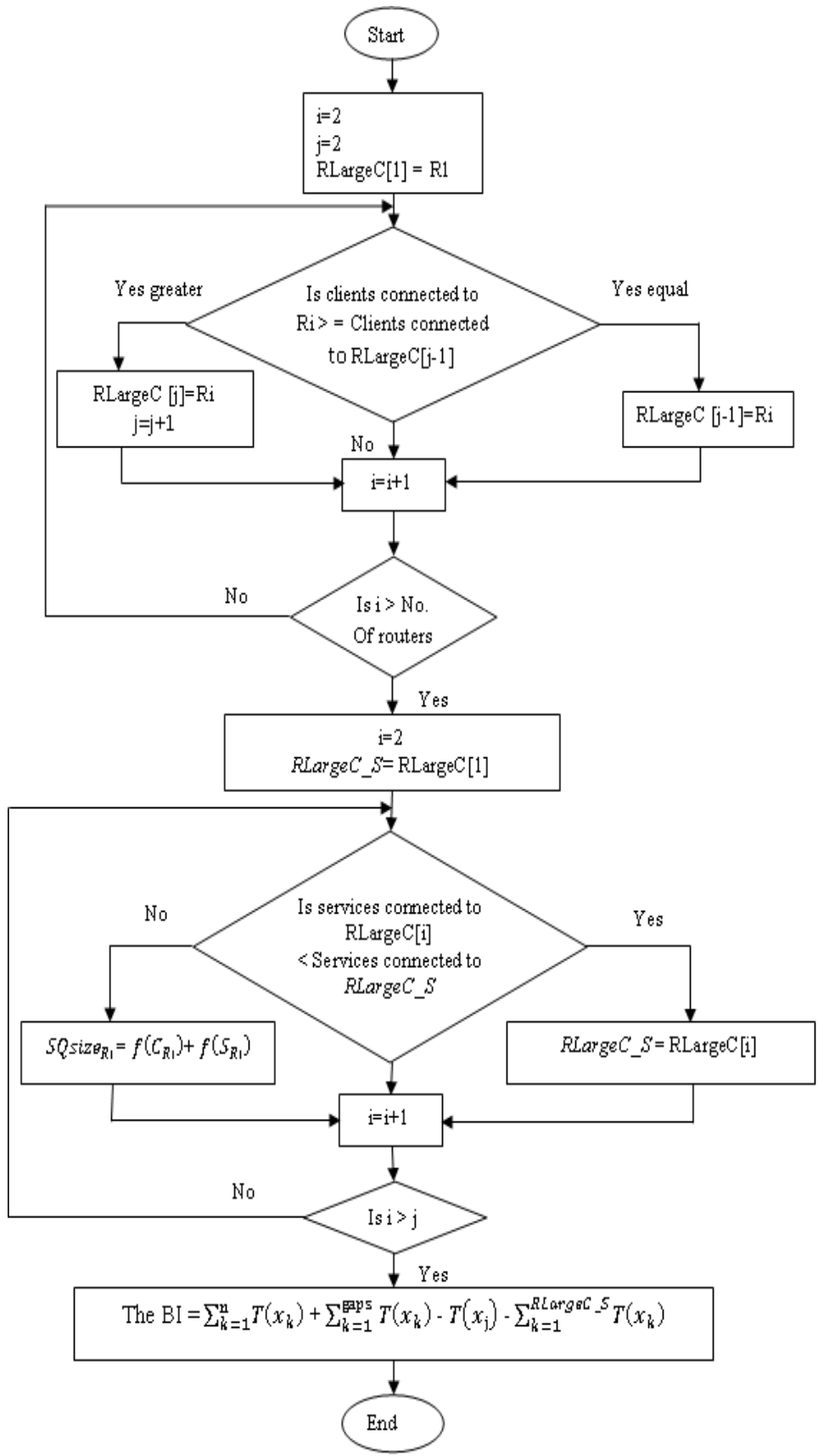

Figure 6. Flowchart of Best interval Algorithm

### 3.2.3 Choosing candidate router rule:

The following rule should be followed to choose the proper router that influences the length of the interval.

1. Identify the router that connected to the largest number of clients and lowest number of services.

## 4. SIMULATION RESULTS

The applied simulated models clarify the influence of increasing the proportion of consumed network resources and the number of clients on the performing of UPnP. It has been compared between the algorithm and normal cases over UPnP. The network design includes 4 routers (R0, R1, R2, and R3) connected in decentralized manner in first model and in centralized manner in second model where each router is connected to 3 services (S0, S1..., and S9) except R2 which connected to six clients (C0, C1... C5). Network parameters have been shown in table 1.

Table1. Network Parameters

| Parameter name | Value |
| --- | --- |
| Bandwidth among routers (Main links) | 512Kb |
| Bandwidth between routers and other nodes (Sub links) | 256Kb |
| Delay in main and sub | 0ms |
| Queue Type | Drop Tail |
| Routing Protocol | DSDV |
| Message Length of discovery (Multicast) | 64 bytes |
| Message Length of discovery reply (Unicast) | 128 bytes |
| Message Length of cross traffic | 100, 200, 300 bytes |
| Simulation Time | 100.0 seconds |

The applied scenario is:

1. (C0, C1... C5) send multicast messages to discover all the services in the network, then,

2. All services send reply messages to the requested clients. In algorithm each service separates any consecutive replying messages with a specific period of time. While, in normal case service replies to the discovery requested dependently.

3. There is a UDP cross traffic (S0 with S8) and (S1 with S7), where S0, S1 are connected to R0 and S7 & S8 are connected to R3. The rate of the cross traffic is 0.01 and the messages size is different.

4. In all tests the algorithm used the same interval regardless the cross traffic

### 4.1 Decentralized Model

In this model the router is connected in Decentralized mode. Fs (7, 8, 9, 10, 11 and 12) explain the impact of increasing the consumption of network resources such as bandwidth on the performing of suggested algorithm and normal case over UPnP. This has been achieved by increasing the size of cross traffic messages.

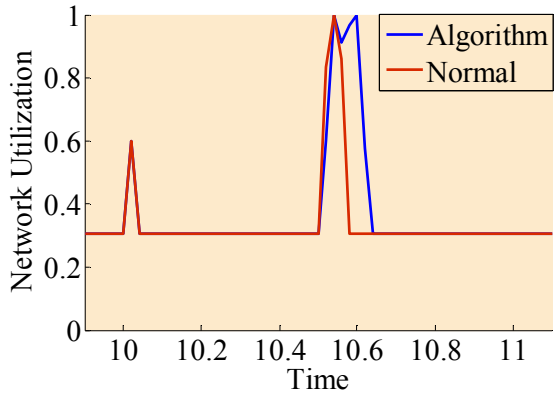

Figure 7 Main links Utilization when cross traffic is 30%.

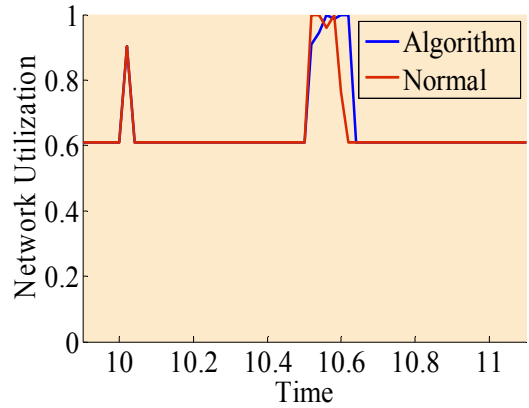

Figure 8 Main links Utilization when cross traffic is 62%.

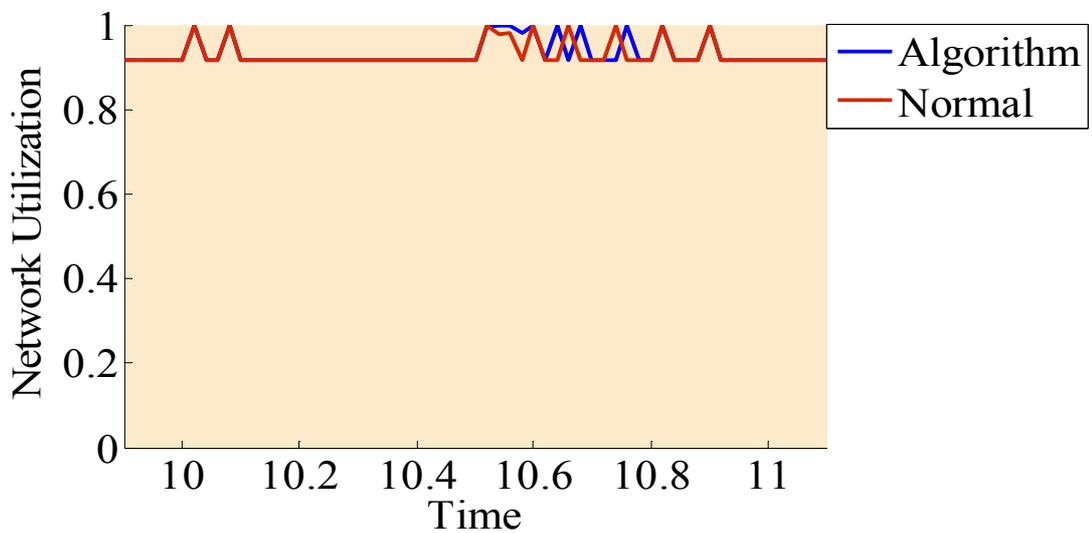

Figure 9 Main links Utilization when cross traffic is 92%.

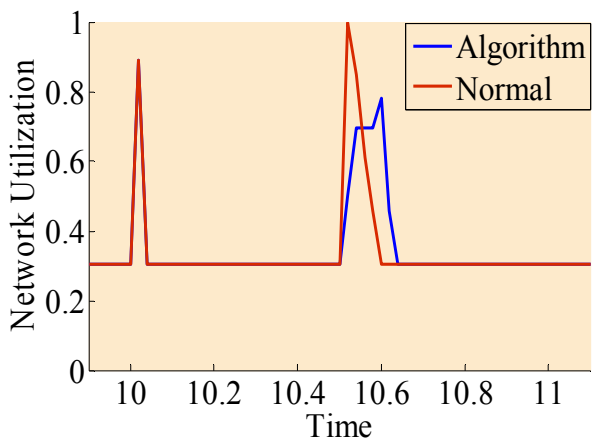

Figure 10 Sub links Utilization when cross traffic is 30%.

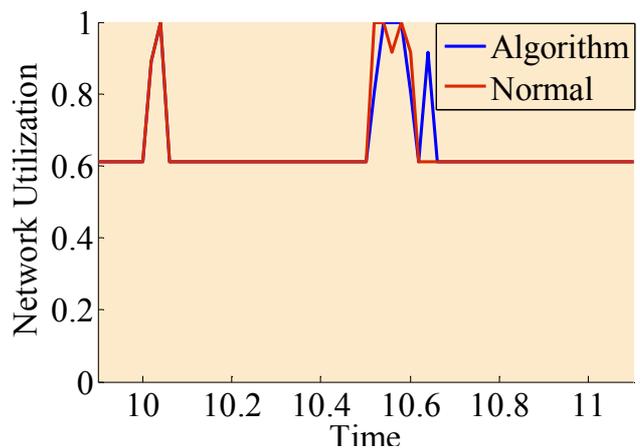

Figure 11 Sub links Utilization when cross traffic is 62%.

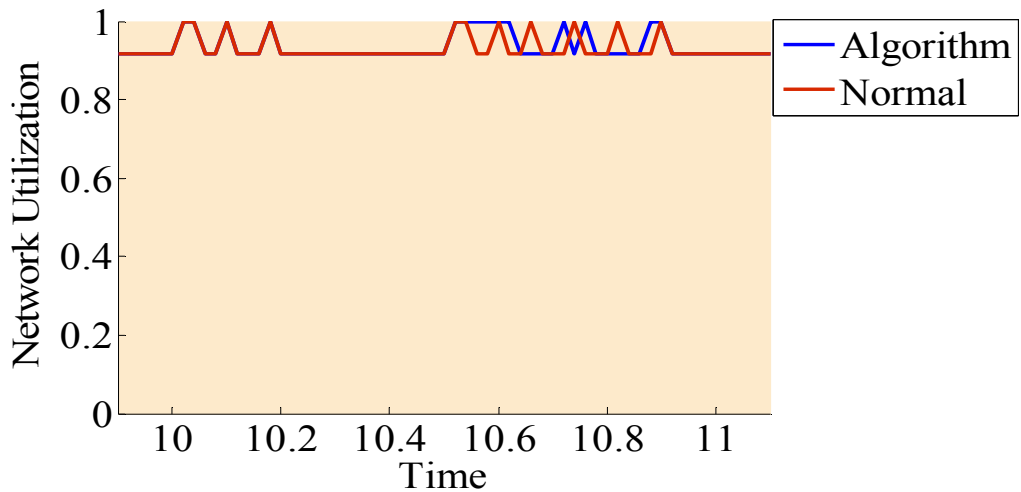

Figure 12 Sub links Utilization when cross traffic is 92%.

The network utilization peaks in Fs (7, 8…, and 12) can be explained by their causes. The cause of first peak(s) which start at $10^{th}$ second of simulation time is the clients' discovery requests' messages and the second peak(s) which start at $10.5^{th}$ second of simulation time is the services replying messages. The waiting interval in algorithm gives more time for the routers to submit the available messages to their destination and this is clear in the previous Fs (in second peak(s)).

As the cross traffic increasing, the available bandwidth decreasing results in more time required to deliver the messages. When the messages incoming rate is more than the available of link capacity the messages will be dropped caused reduction in network utilization after it has been reached the maximum usage. This is attributed to the routing management mechanism too, which it could produce bursty losses during congestions and high delays (by dividing the sending rate into half) [12]. Consequently, the network utilization is reduced when it is reach the maximum usage.

F 13 and 14 represent the discovery rate and the dropping rate in algorithm and normal case. Where the dropping rate is calculated based on the number of sent messages during the period of sending the services' reply messages including the backward messages.

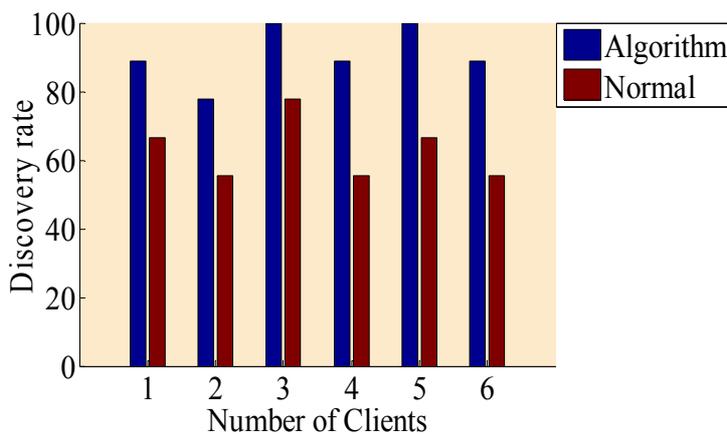

Figure 13 discovery rates for each client when cross traffic is 30%.

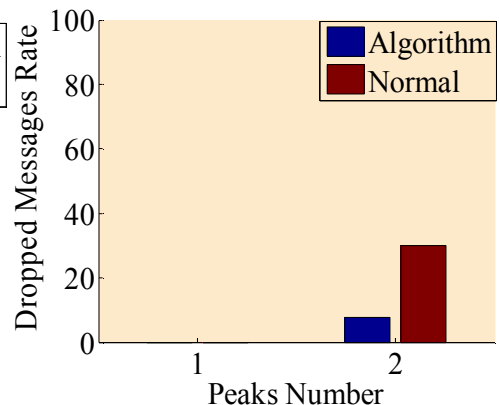

Figure 14 Dropping rate in two when cross traffic is 30%.

In algorithm case the discovery rate was ranging between 78% and 100%, while in normal case it was between 56% and 78%. There were not dropped messages in both algorithm and normal case in clients discovery phase, but there were dropped messages in services' replying messages phase. The dropping rate in normal case was more than four times as dropping rate in algorithm. The algorithm guarantees reasonably high discovery rate for all the involved clients and less dropped messages. The implementations show that if the number of clients is increased to be 8 clients the discovery rate would be in the same range in algorithm, but it would reduce in normal case to be between (45% and 78%). On the other hand the number of dropped messages will increase in both cases, but it is relatively increasing in algorithm case.

## 4.2 Centralized model

In this model additional router (R4) is add to the topology of the suggested network to acts as the central node that all the other routers should connect to it. Fs (15, 16, 17, 18, 19 and 20) explain the impact of increasing the consumption of network resources such as bandwidth on the performing of suggested algorithm and normal case over UPnP.

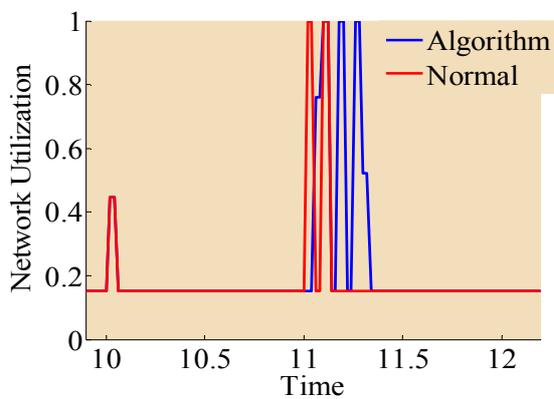

Figure 15 Main links Utilization when cross traffic is 15%.

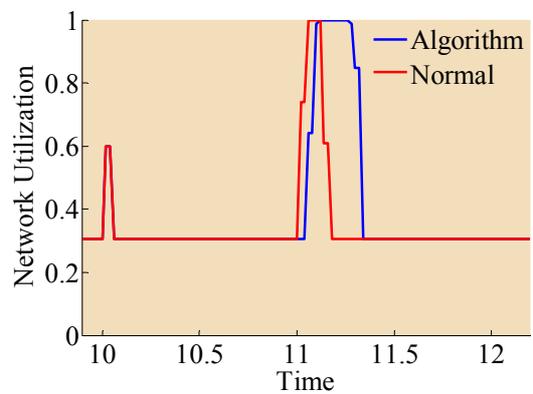

Figure 16 Main links Utilization when cross traffic is 30%.

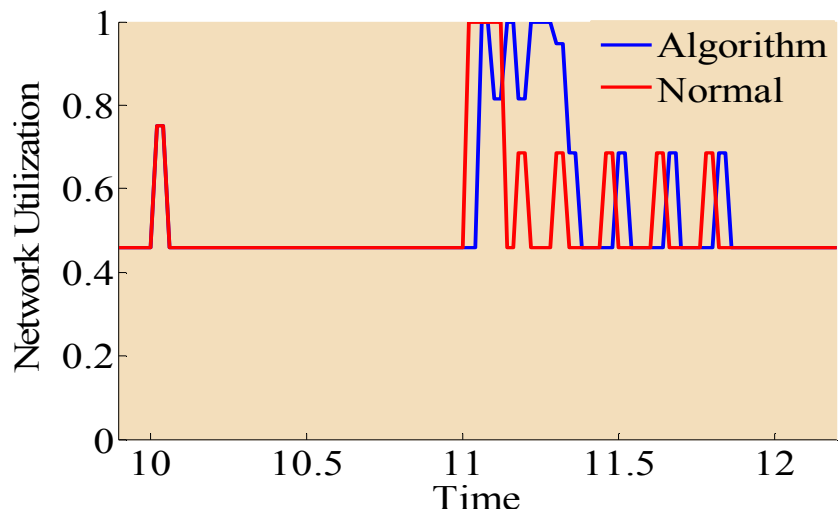

Figure 17 Main links Utilization when cross traffic is 45%.

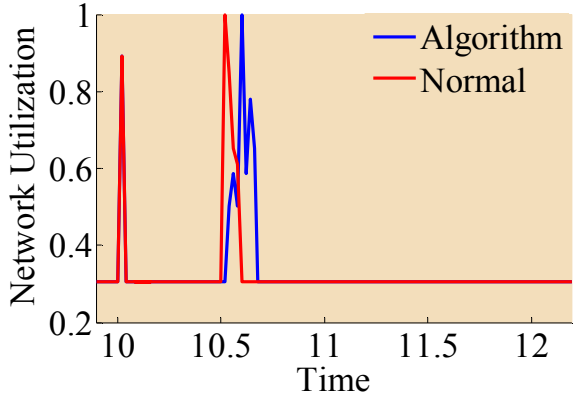

Figure 18 Sub links Utilization when cross traffic is 30%.

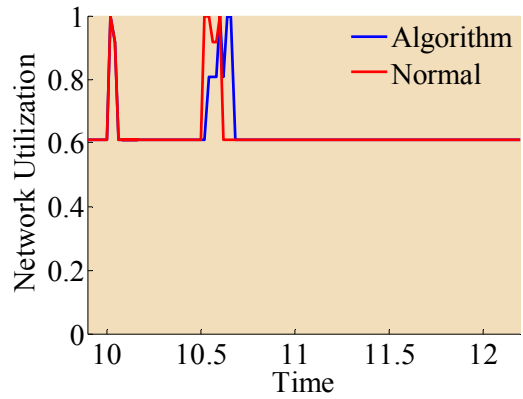

Figure 19 Sub links Utilization when cross traffic is 62%.

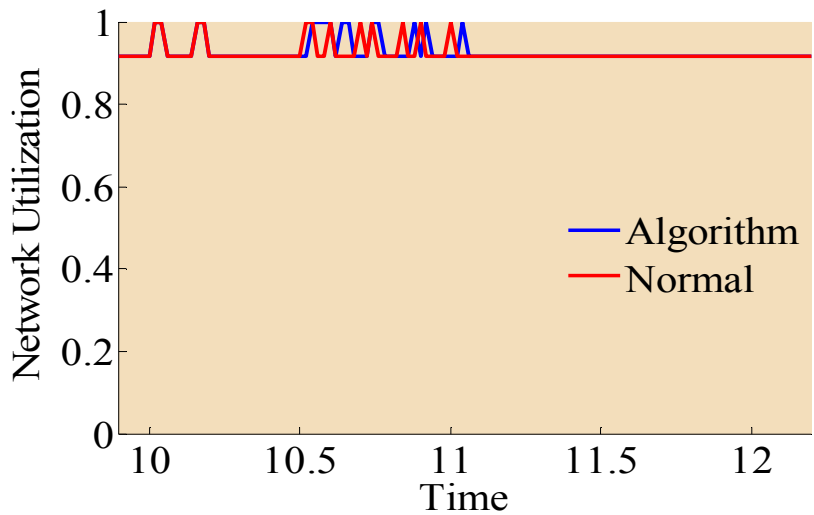

Figure 20 Sub links Utilization when cross traffic is 92%.

Although the used cross traffic is the same as the used in the first model, the network utilization is divided into half as explained in Fs (15, 16… and 20). This is attributed to the path of the messages which is affected by the network topology.

The causes of the network utilization peaks in Fs (15, 16…, and 20) are the same causes as in the first model. The available bandwidth decrease when the cross traffic increases. This result in, increase the consumed time period to deliver the messages.

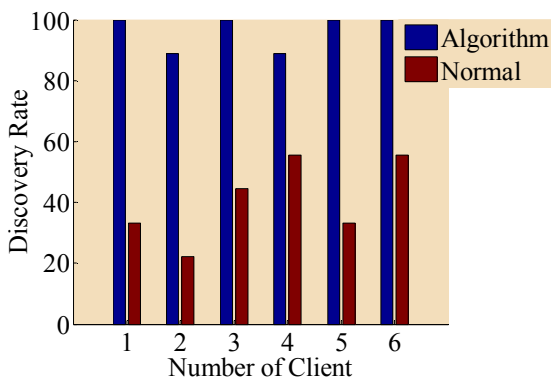

Figure 21 discovery rates for each client when cross traffic is 15%.

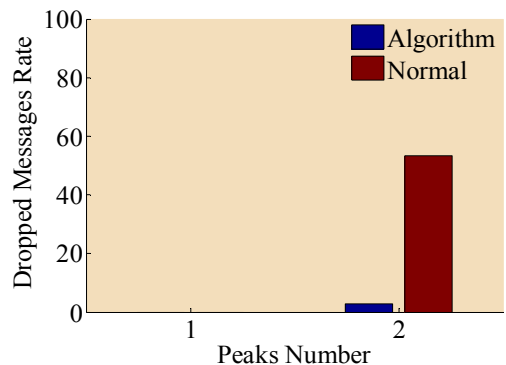

Figure 22 Dropping rate in two when cross traffic is 15%.

F 21 and 22 shows the discovery rate and dropping rate respectively. In algorithm case the discovery rate between 88.88% and 100%, while in normal case it was between 22.22% and 55.55%. The discovery rate in normal case is very limited and this is because the topology of the network where the root router represents the bottleneck in the path of any message. The same cause is behind the big difference between the dropped messages in algorithm and normal cases.

## 5. CONCLUSION

The aims of the proposed algorithm is to use the maximum of the available network resources and achieve optimal results by avoid dropping messages and speedily in performing the discovery process in services discovery protocols. The results show the delay in sending services response messages produce smoothing and speedily in discovery procedure. In addition they clarify improving in discovery rate and in dropping rate when the algorithm is employed in both configuration (Decentralized and Centralized). Further, the increasing in the number of clients does not affects the performance of the suggested algorithm, but it is influenced by the proportion of consumed network resources and this reveal the need for providing additional/different parameters in the suggested algorithm to cope with the changing in the available network resources. The network topology impacts the best interval and the queue size calculations and this explicates the need to conclude a general algorithm which can be employed with variety topologies.

### REFERENCES

[1]   Jim Edwards and Stephen Murtha, "DLNA Home Networked Interoperability Guidelines", Expanded 2006.

[2]   Al-Mejibli, I and Colley, M, "Evaluating Transmission Time of Service Discovery Protocols by using NS2 Simulator", Wireless Advanced (WiAD), 2010 6th Conference on London.

[3]   Members of the UPnP Forum, "UPnP™ Device Architecture 1.1", 2008.

[4]   Christopher N. Ververidis and George C. Polyzos, "Service Discovery for Mobile Ad Hoc Networks: A Survey of Issues and Techniques", "http://mm.aueb.gr/publications/2008-SD-SURVEY-COMST.pdf.", 14 November 2006.

[5]   Michel Barbeau, Evangelos Kranakis and Honghui Luo, "Strategies for Service Discovery over Ad Hoc Networks", "http://www.engineeringletters.com/issues_v13/issue_1/EL_13_1_2.p df", 4 May 2006.

[6]   "Novell Documentation: Novell eDirectory 8.7 - How SLP Works", "http://www.novell.com/documentation/edir87/?page=/documentation /edir87/edir87/data/a60jiyy.html".

[7]   Eugene A. Gryazin, "Service Discovery in Bluetooth",Helsinki University of Technology.

[8]   Chris Custine, "Introduction to OSGi", Denver Java User Group - November 12, 2008.

[9]   OSGi Alliance, "About the OSGi Service Platform", Technical Whitepaper, Revision 4.1, 7 June 2007.

[10]  Kevin Mills and Christopher Dabrowski, "Adaptive Jitter Control for UPnP M-Search", ICC '03. IEEE International Conference, May 2003.

[11]  Kevin Mills and Christopher Dabrowski, "Adaptive Jitter Control for UPnP M-Search", ICC '03. IEEE International Conference, May 2003.

[12]  Sneha K. Kasera, Ramachandran Ramjee, Sandra Thuel and Xin Wang, "Congestion Control Policies for IP-based CDMA Radio Access Networks", Holmdel, New Jersey, 2005.